\documentclass[12pt]{article}
\usepackage[margin=1in]{geometry}
\usepackage{booktabs,multirow,graphicx,amsmath,amssymb,amsfonts}
\usepackage{caption}
\usepackage{subcaption}
\usepackage{float}
\usepackage{url}
\usepackage[hidelinks]{hyperref}
\title{Network-distance decay of perceived online social support}
\author{Masanori Takano$^{1,2,*}$, Kenji Yokotani$^{3}$,\\
Masaki Chujyo$^{4}$ and Fujio Toriumi$^{4}$\\[0.75em]
\small $^{1}$Multidisciplinary Information Science Center, CyberAgent, Inc., Tokyo, Japan\\
\small $^{2}$Center of Advanced Research for Human-AI Symbiosis Society,\\
\small Keio University, Tokyo, Japan\\
\small $^{3}$Graduate School of Technology, Industrial and Social Sciences,\\
\small Tokushima University, Tokushima, Japan\\
\small $^{4}$Graduate School of Engineering, The University of Tokyo, Tokyo, Japan\\
\small $^{*}$Correspondence: takano\_masanori@cyberagent.co.jp}
\date{}

\begin{document}
\maketitle

\begin{abstract}
Perceived social support can buffer stress, but how it is associated across online social networks at different graph distances remains unclear. Here we show that inferred perceived online social support in a large avatar communication application decays with network distance in a form better described, over the observed range, by a power-decay model than by a single exponential. We linked two-wave survey data from Pigg Party with behavioral logs, trained a random forest to infer perceived support for active users, and regressed Wave 2 scores on Wave 1 scores for users at hop distance $k$, adjusting for baseline support and covariates.
Adjusted associations persisted across hops, consistent with power-law-like decay over the observed range.
Individual-based simulations indicated that heterogeneous source-specific exponential decay rates can generate heavy-tailed aggregate decay. These results suggest that network position heterogeneity should be considered when characterizing distance-dependent associations among psychosocial states in online communities.
\end{abstract}

\section{Introduction}

Many collective phenomena in social networks are studied by assigning states to nodes and asking how these states are associated, transmitted, or correlated across edges~\cite{Castellano2009,Jusup2022,Caldarelli2026}. Epidemic processes, information diffusion, and cooperative behavior provide standard examples in which local interactions generate macroscopic patterns over network distance \cite{PastorSatorras2015,Guille2013,Zhang2016,Fowler2010}. A central question in such systems is the shape of distance decay: whether an association is essentially local, with a single characteristic length scale, or whether it persists across broader network neighborhoods. While this question has been extensively examined for infections, information, and behavioral cascades, it remains less clear how psychological states measured at the individual level are organized over graph distance in large online social networks.

Here, we focus on perceived online social support as such a node-level psychological state.
Perceived social support represents an individual's perception that sufficient social resources are available to provide emotional comfort, practical assistance, and guidance during challenging circumstances.
This can buffer the harmful effects on mental health among people facing difficulties \cite{Che2018,Rothon2011}, a phenomenon often referred to as the stress-buffering effect \cite{Cohen1985}. Online communities are especially relevant for studying this state because they can provide opportunities for support for people who lack social resources in the physical world, while reducing interpersonal risk and enabling flexible self-disclosure \cite{takano_icwsm2022,takano_icwsm2025,takano_chbr2026}. Avatar communication is a particularly informative setting: users interact through virtual bodies, facial expressions, and gestures in shared virtual spaces, enabling nonverbal, real-time interaction with online co-presence \cite{MasanoriTakano2019,takano_icwsm2025}.

Digital platforms also enable the connection between psychological measurement and network structure. In offline settings, it is difficult to observe, at scale, who interacted with whom and for how long. By contrast, online communities record visits, replies, co-presence, follow relationships, and other interaction events, from which large-scale user networks can be constructed. Questionnaires provide node-level measurements of psychological states, whereas behavioral logs provide the corresponding interaction network. Combining these two data sources enables the treatment of perceived online social support as a measurable state variable assigned to nodes in a social network. Related network approaches have examined information exchange and social support in web-based health communities \cite{Liu2020}, and, more generally, digital traces can provide behavioral markers from which psychological traits or states may be inferred \cite{Stachl2020,Yokotani2025_socanx}.

This perspective differs from much of the existing statistical physics and complex network literature on social support. In epidemic-process models, for example, social support is often represented as a local resource or transition-rate modifier: healthy neighboring nodes allocate support resources to infected nodes, thereby changing recovery probabilities and macroscopic outcomes such as epidemic thresholds, phase transitions, or hysteresis \cite{PastorSatorras2015,Chen2018SS,Zhou2019}. This line of work demonstrates that support-like local interactions can alter the dynamics of spreading. However, it usually abstracts social support as a resource-allocation mechanism operating through adjacency, rather than as a psychological perception held by users. The complementary empirical question addressed here is whether one user's perceived support score is associated not only with the later support scores of directly connected users, but also with those of users at longer graph distances.

Such distance-conditioned associations may arise through several mechanisms. Supportive experiences may influence a recipient's psychological state and alter the probability that the recipient subsequently provides support to another person. These chains are related to emotional-support cascades \cite{Lakon2017}, cooperative behavior cascades \cite{Fowler2010}, and pay-it-forward or upstream reciprocity \cite{Nowak2006b,Iwagami2010,Horita2016,Obayashi2023}. The present study does not assume a pathogen-like contagion process, nor does it identify explicit sequences of support provision. Instead, it quantifies longitudinal associations among perceived-support states as a function of shortest-path distance in an interaction network.

The functional form of this distance decay is informative. In homogeneous media or linear spreading processes with a single characteristic length scale, effects often decay exponentially with distance. Under such a description, distant nodes contribute little, and the association is mainly local. Real social networks, however, are structurally and dynamically heterogeneous. Perceived support in online communities plausibly depends not only on the existence of contact but also on interaction frequency, replies, visits, co-presence, conversation continuity, tie strength, and message content. Weighted social networks and egocentric communication networks show broad heterogeneity in interaction weights, communication rates, and tie strengths \cite{Barrat2004,Onnela2007,takano2018_rsos,Iniguez2023,Heydari2024}. Edge weights and bursty human activity can also slow spreading even in networks with short topological paths \cite{Karsai2011,Iribarren2009,Min2011}. These observations suggest that perceived-support associations may decay more slowly than a single exponential over network distance.

One mechanism for such slow aggregate decay is heterogeneity in source-specific decay rates. A source located near hubs, branching regions, or shortcut-rich parts of the network may have many alternative routes and reachable nodes, leading to a slowly decaying distance profile. A source near the periphery may have a faster decay rate. Even if an exponential decay locally approximates the association from each source, variation in these source-specific decay rates can produce an aggregate pattern that is heavier-tailed than a single exponential \cite{Beck2003,Mitzenmacher2003}. A mixed exponential distribution provides a simple example in which heterogeneity in exponential rates yields a slowly decaying aggregate form \cite{Keatinge1999}. In this study, we use heavy-tailed decay to denote decay slower than a single exponential over the observed distance range, and we operationalize a power-decay model as a linear relation between $\log \beta_k$ and $\log k$.

We analyze Pigg Party, a large Japanese avatar communication application, by combining behavioral log-based inference, empirical network scaling analysis, and individual-based modeling. First, we link two-wave survey data with behavioral logs and train a prediction model to infer perceived online social support for active users who did not respond to the survey. Second, we examine whether the Wave 1 support score of a source user is associated with the Wave 2 support score of target users at hop distance $k$ in a room-visit interaction network, while adjusting for the target user's baseline support score and covariates. Third, we develop an individual-based model to test whether heterogeneous source-specific exponential decay rates can generate heavy-tailed aggregate distance decay under a common local updating rule. Together, these analyses characterize perceived online social support as a networked psychological state and provide a mechanism by which distance-dependent associations among psychosocial states can extend beyond immediately adjacent users.
\begin{figure}[H]
\centering
\includegraphics[width=0.55\textwidth]{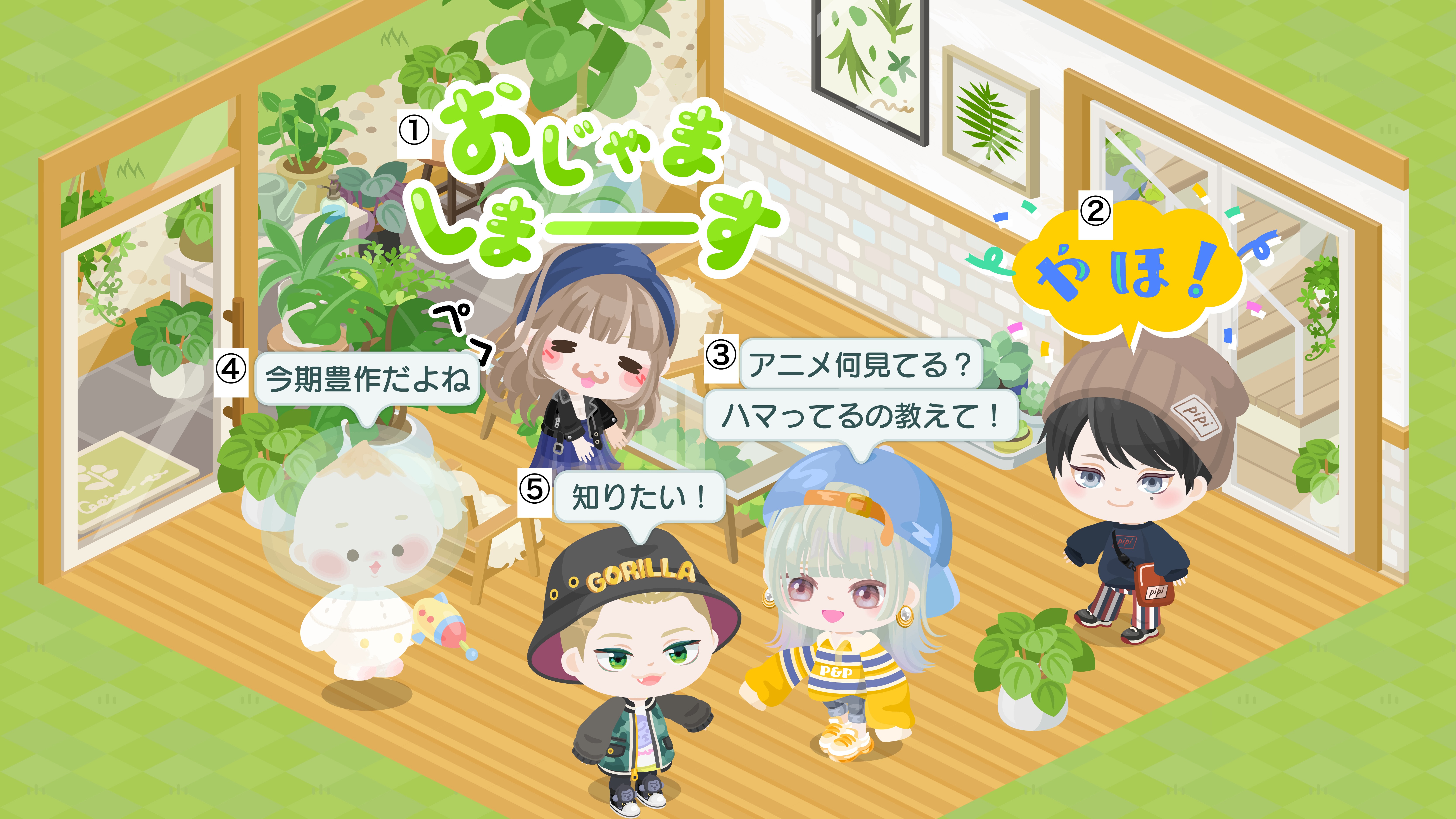}
\caption{\textbf{Pigg Party avatar communication space.} Players communicate synchronously through avatars in virtual rooms. The image illustrates the type of avatar-mediated, real-time interaction environment with online co-presence from which the behavioral logs were obtained.}
\label{fig_piggparty}
\end{figure}

\section{Results}

\subsection{Inferring perceived online social support from behavioral logs}

We evaluated whether perceived online social support could be inferred from behavioral and profile features. The prediction dataset consisted of 3,668 wave-level observations. To prevent leakage across waves for the same user, we split the data by user into training (80\%), validation (10\%), and test (10\%) sets. Four supervised learning algorithms were compared: random forest, LightGBM, Elastic Net regression, and a linear-kernel support vector machine.

After hyperparameter tuning, we compared the four algorithms on the user-separated test set. The random forest was selected because it achieved the highest correlation between predicted and observed support ($r=0.362$) and had the lowest root mean squared error and the highest $R^2$ (Table~\ref{tab:perf}). LightGBM had the lowest mean absolute error (MAE). The correlation for the random forest is comparable to previously reported correlations for predicting psychological measures from digital traces~\cite{Stachl2020}.
The selected random forest used 1000 trees, a node size of 30, and 30 randomly sampled candidate variables at each split. We therefore used this model to infer perceived online social support for active users who did not respond to the survey.

\begin{table}[H]
\centering
\caption{\textbf{Comparative prediction performance for perceived online social support.} The four tuned algorithms were compared on a test set containing users who did not appear in the training or validation sets. Corr. denotes the correlation between predicted and observed scores.}
\label{tab:perf}
\small
\begin{tabular}{lrrrr}
\toprule
Model & MAE & RMSE & $R^2$ & Corr. \\
\midrule
Random forest & 0.878 & {\textbf 1.106} & {\textbf 0.122} & {\textbf 0.362} \\
LightGBM & {\textbf 0.866} & 1.139 & 0.068 & 0.349 \\
Elastic Net & 0.880 & 1.115 & 0.108 & 0.341 \\
SVM, linear & 0.906 & 1.214 & -0.058 & 0.201 \\
\bottomrule
\end{tabular}
\end{table}

The test-set correlation of $r=0.362$ exceeds the conventional medium benchmark of $r=0.30$~\cite{Cohen1992}, while the test-set $R^2$ indicates that the model explained 12.2\% of the variance in questionnaire-based perceived support among users not included in training or validation. Because such benchmarks are context-dependent and most individual-level variance remains unexplained, we characterize the performance as moderate rather than high. The inferred scores are treated as noisy, incomplete proxies for aggregate network-level analysis, not as substitutes for questionnaire scores in individual-level assessment or decision-making.

\subsection{Network-distance decay in the Pigg Party interaction network}

We examined whether inferred perceived online social support showed a distance-dependent longitudinal association over the Pigg Party interaction network. The coefficient $\beta_k$ quantifies the adjusted association between the Wave 1 support score of a source user at hop distance $k$ and the Wave 2 support score of the target user, after controlling for the target user's Wave 1 support score and covariates. The distance $d(i,j)=k$ was defined as the shortest-path distance in the Wave 1 interaction network.

\begin{figure}[H]
\centering
\begin{subfigure}[b]{0.47\textwidth}
\centering
\includegraphics[width=\textwidth]{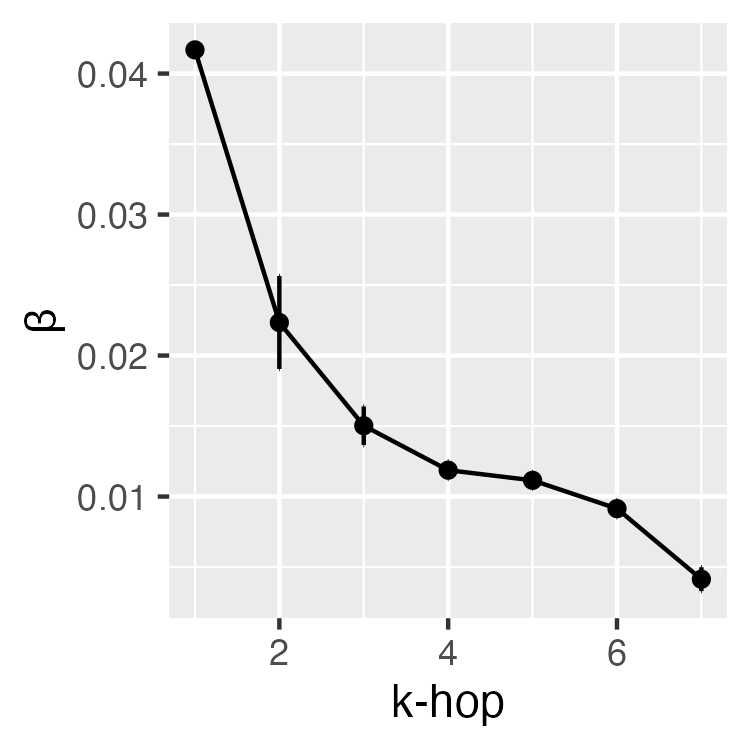}
\caption{Linear scale}
\end{subfigure}
\hfill
\begin{subfigure}[b]{0.47\textwidth}
\centering
\includegraphics[width=\textwidth]{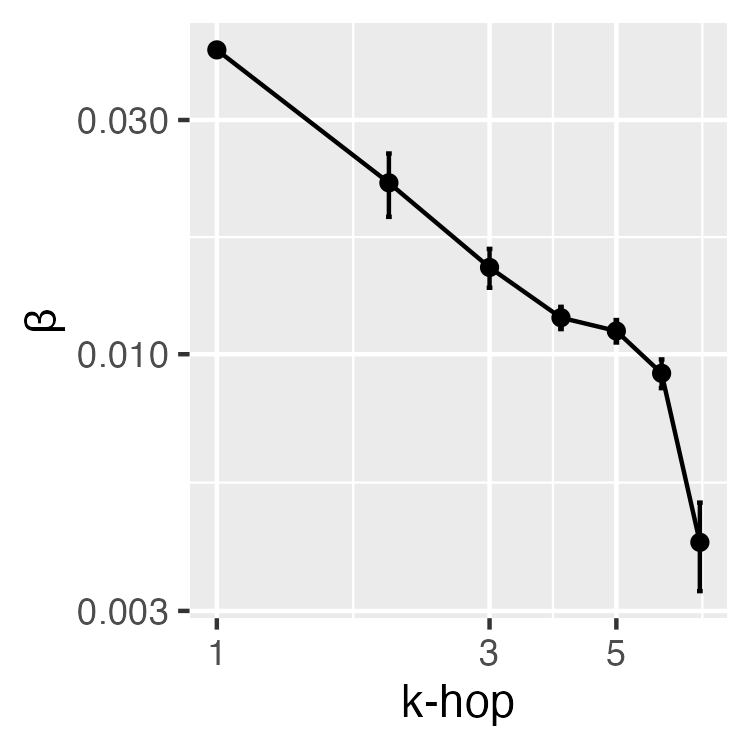}
\caption{Log-log scale}
\end{subfigure}
\caption{\textbf{Network-distance decay of inferred perceived online social support.} Points show the mean coefficient $\beta_k$ across 100 sampling repetitions. Error bars show standard deviations across repetitions, not confidence intervals for independent pairs.}
\label{fig_dis}
\end{figure}

The adjusted association was positive over several hops and decreased with hop distance (Fig.~\ref{fig_dis}). Over the range $1\leq k\leq6$, the decline was better described by the power-decay model than by the single-exponential model. Specifically, in the Davidson-MacKinnon $J$ test, adding the fitted values from the power-decay model to the exponential model significantly improved the fit, whereas adding the fitted values from the exponential model to the power-decay model did not (Table~\ref{tbl_pg_model_test}).

The prediction model included features representing local interaction environments, such as the number of mutual follows, the degree, and the mean edge weight. These features are theoretically relevant because perceived online social support is expected to depend on the frequency of local contact and relationship strength. At the same time, both inferred support scores and network distance are partly derived from interaction logs, which could strengthen the association between adjacent nodes ($k=1$). We therefore repeated the comparison after excluding $k=1$ and analyzing only $2\leq k\leq6$. The power-decay model again showed a higher adjusted $R^2$ than the exponential model (0.973 versus 0.905), and the $J$ test provided marginal evidence against the exponential model while not rejecting the power-decay model (Supplementary Note~1). This suggests that the heavy-tailed pattern was not driven solely by adjacent-node associations.

\begin{table}[H]
\centering
\caption{\textbf{Descriptive comparison of distance-decay models in Pigg Party.} Model comparison was performed over $1\leq k\leq6$. $M_e$ is the single-exponential model and $M_p$ is the power-decay model.}
\label{tbl_pg_model_test}
\small
\begin{tabular}{llrrrr}
\toprule
Null model tested & Adj. $R^2$ & Added fitted & Estimate & S.E. & $P$ value \\
\midrule
$M_e$: $\log\beta_k\sim k$ & 0.870 & fitted($M_p$) & 1.258 & 0.149 & 0.003 \\
$M_p$: $\log\beta_k\sim\log k$ & 0.989 & fitted($M_e$) & -0.281 & 0.156 & 0.170 \\
\bottomrule
\end{tabular}
\end{table}

\subsection{A heterogeneity model for heavy-tailed distance decay}

The empirical analysis showed a decay pattern slower than a single exponential. To interpret this pattern, we considered the possibility that exponential decays with different scales are superposed~\cite{Beck2003}. Let $\beta_{k\mid s}$ denote the effect of source $s$ on nodes at distance $k$. Conditional on a fixed source, we approximate the local decay by
\begin{equation}
\beta_{k\mid s}\approx A_s\exp(-\kappa_s k),
\label{eq:source_decay}
\end{equation}
where $A_s$ represents the source-specific magnitude and $\kappa_s$ represents the source-specific decay rate. $\kappa_s$ is an observed summary of how rapidly the influence of source $s$ decays under the same microscopic updating rule. A source located near hubs or branching regions can have many alternative routes and reachable nodes, so that its effect may decay slowly with distance (small $\kappa_s$). A source near the periphery of the network may have a faster decay rate (large $\kappa_s$).

The empirical regression coefficient $\beta_k$ is aggregated across sources. Thus, it can be approximated as
\begin{equation}
\beta_k\approx \mathbb{E}_s\left[A_s\exp(-\kappa_s k)\right].
\label{eq:mixture_decay}
\end{equation}
A mixture of exponential functions can generate decay slower than a single exponential and can appear close to a power law over a finite range~\cite{Mitzenmacher2003}. For example, a mixed exponential model in which $\kappa$ follows a gamma distribution yields a form proportional to $(1+\theta k)^{-\alpha}$~\cite{Keatinge1999}. Thus, an apparent power-decay pattern in $\beta_k$ does not require a microscopic propagation rule that is intrinsically power-law-like. It can emerge from the aggregation of source-specific exponential decays, whose rates are widely distributed due to network heterogeneity.

\subsection{Simulations on model-generated networks}

To examine whether the heterogeneity mechanism described above can arise from network structure under a common local rule, we used an individual-based, event-driven support-updating model. In this model, each node carries a perceived-support state, and a random-walk-like process with reply and restart probabilities generates message events. When a node receives a message, only the receiver's state is updated by combining it with input proportional to the sender's state. For each source, we introduced a small perturbation to its initial state. We measured the resulting change in the final states of nodes at distance \(k\), yielding the source-specific profile \(\beta_{k\mid s}\); averaging these profiles over sources gave the aggregate \(\beta_k\). Thus, source-specific decay rates \(\kappa_s\) were estimated from simulated distance profiles rather than imposed as parameters.

\begin{figure}[H]
\centering
\begin{subfigure}[b]{0.31\textwidth}
\centering
\includegraphics[width=\textwidth]{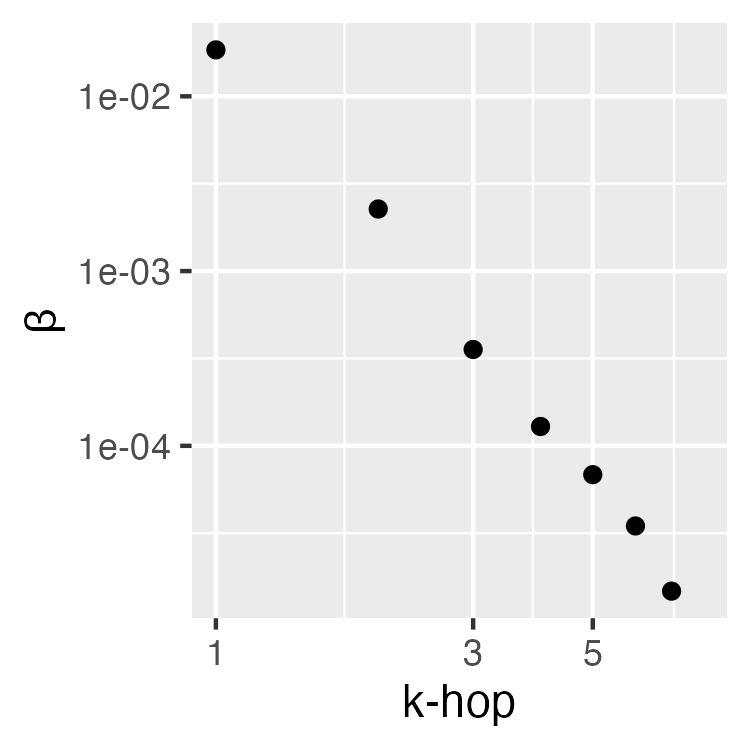}
\caption{BA model}
\end{subfigure}
\hfill
\begin{subfigure}[b]{0.31\textwidth}
\centering
\includegraphics[width=\textwidth]{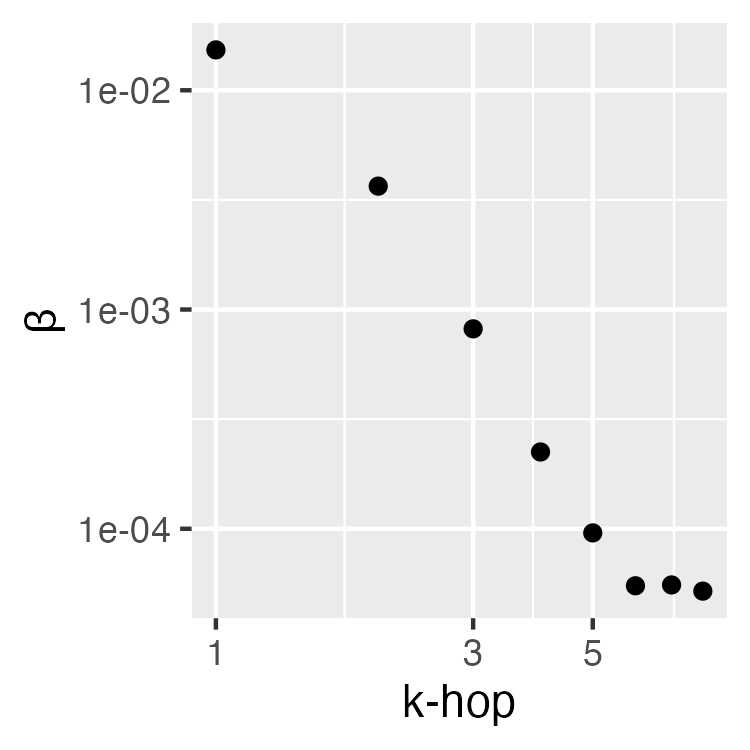}
\caption{ER model}
\end{subfigure}
\hfill
\begin{subfigure}[b]{0.31\textwidth}
\centering
\includegraphics[width=\textwidth]{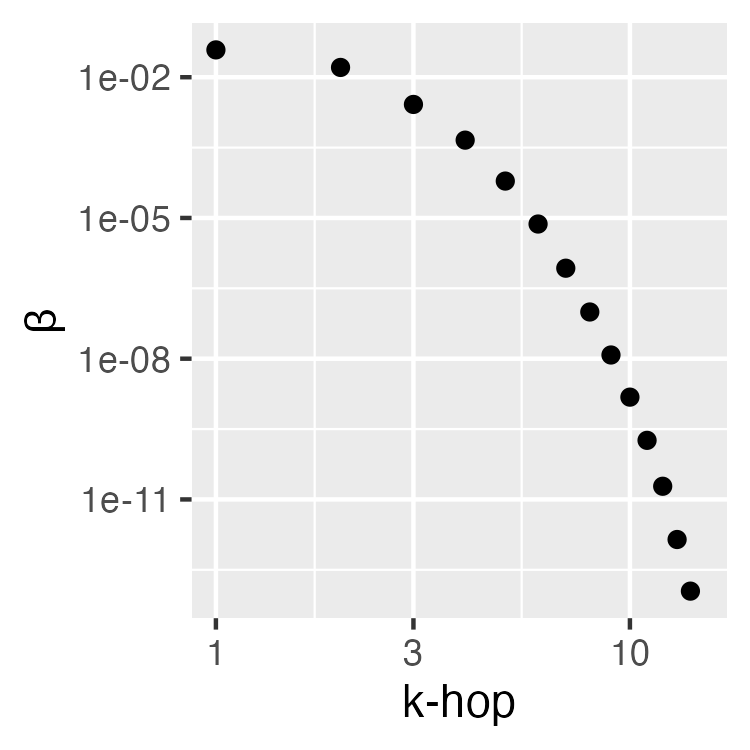}
\caption{WS model}
\end{subfigure}
\caption{\textbf{Simulated distance-decay curves on model-generated networks.} Points show the simulated coefficient $\beta_k$ as a function of hop distance on log-log axes.}
\label{fig_sim_res}
\end{figure}

We applied the same local updating rule to three model-generated network structures: scale-free Barab\'asi--Albert (BA)~\cite{Barabasi1999}, Erd\H{o}s--R\'enyi (ER) random~\cite{Erdos1959}, and Watts--Strogatz (WS) small-world~\cite{Watts1998} networks. This comparison allowed us to examine whether network topology alone can produce broader or narrower distributions of \(\kappa_s\) and, consequently, different aggregate
distance-decay patterns.
The resulting distance-decay curves differed across network structures (Fig.~\ref{fig_sim_res}). In the BA network, the $J$ test rejected the single-exponential model, whereas the power-decay model was not rejected (Table~\ref{tbl_model_test}). The ER network showed the same direction, although the evidence against the exponential model was marginal. In the WS network, by contrast, the exponential model provided a higher descriptive fit than the power-decay model.

\begin{figure}[H]
\centering
\includegraphics[width=0.6\textwidth]{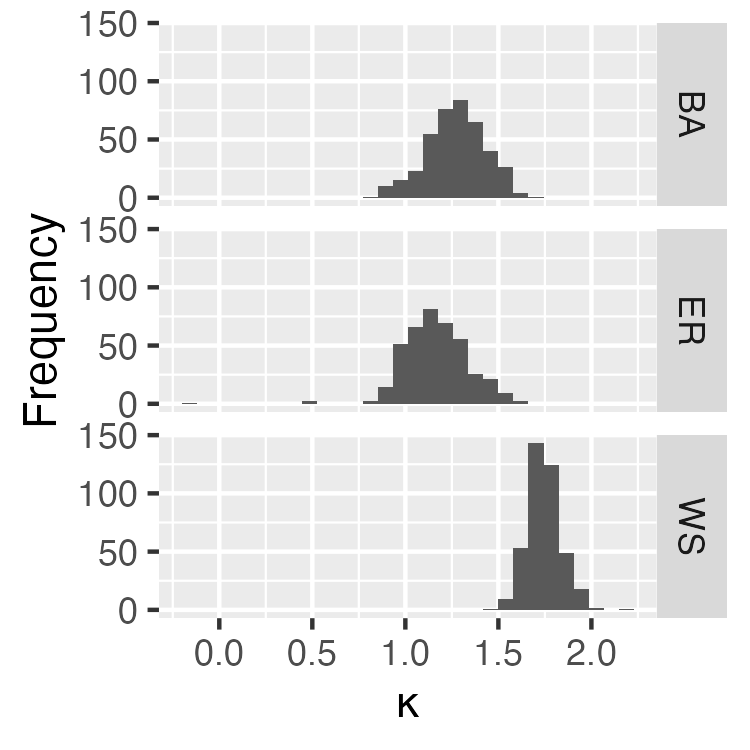}
\caption{\textbf{Distribution of source-specific decay rates on model-generated networks.} The distributions summarize $\kappa_s$ derived from source-level distance profiles using the exponential form in Eq.~\eqref{eq:source_decay}.}
\label{fig_kappa_freq}
\end{figure}

For each source, we summarized the distance profile $\beta_{k\mid s}$ using the exponential form in Eq.~\eqref{eq:source_decay} and examined the distribution of $\kappa_s$ (Fig.~\ref{fig_kappa_freq}). The BA and ER networks had broader $\kappa_s$ distributions than the WS network and contained sources with small $\kappa_s$, corresponding to slowly decaying effects. This source-level heterogeneity provides a mechanism by which aggregate $\beta_k$ can have a heavier tail than a single exponential.

We further examined whether $\kappa_s$ was related to network position. In the BA and ER networks, $\kappa_s$ was negatively correlated with degree, betweenness, closeness, and PageRank (Table~\ref{tbl_kappa_cor}). These correlations indicate that nodes with more outward routes, greater accessibility in random walks, shorter distances to other nodes, or greater brokerage between communities tended to have smaller decay rates. The clustering coefficient, which reflects connections among neighbors, was not strongly associated with $\kappa_s$.

\begin{table}[H]
\centering
\caption{\textbf{Descriptive comparison of simulated decay models on model-generated networks.} The evaluation range was $k\leq6$ for BA and ER networks and $2\leq k\leq14$ for the WS network.}
\label{tbl_model_test}
\small
\begin{tabular}{llllrrr}
\toprule
Network & Null model tested & Adj. $R^2$ & Added fitted & Estimate & S.E. & $P$ value \\
\midrule
BA & $M_e$: $\log\beta_k\sim k$ & 0.927 & fitted($M_p$) & 0.889 & 0.113 & 0.0014 \\
BA & $M_p$: $\log\beta_k\sim\log k$ & 0.994 & fitted($M_e$) & 0.117 & 0.116 & 0.363 \\
ER & $M_e$: $\log\beta_k\sim k$ & 0.966 & fitted($M_p$) & 0.552 & 0.211 & 0.0792 \\
ER & $M_p$: $\log\beta_k\sim\log k$ & 0.973 & fitted($M_e$) & 0.464 & 0.212 & 0.116 \\
WS & $M_e$: $\log\beta_k\sim k$ & 0.997 & fitted($M_p$) & -0.043 & 0.061 & 0.500 \\
WS & $M_p$: $\log\beta_k\sim\log k$ & 0.919 & fitted($M_e$) & 1.040 & 0.059 & $<0.001$ \\
\bottomrule
\end{tabular}
\end{table}

\begin{table}[H]
\centering
\caption{\textbf{Correlations between $\kappa_s$ and node-level network statistics in model-generated networks.} Values in parentheses are $P$ values.
We did not report centrality correlations for the WS network because its near-regular structure makes these centrality measures less informative.}
\label{tbl_kappa_cor}
\small
\begin{tabular}{lrrrrr}
\toprule
Model & Degree & Clustering & Betweenness & Closeness & PageRank \\
\midrule
BA & -0.62 ($<0.001$) & -0.04 (0.482) & -0.60 ($<0.001$) & -0.73 ($<0.001$) & -0.59 ($<0.001$) \\
ER & -0.50 ($<0.001$) & 0.02 (0.659) & -0.56 ($<0.001$) & -0.51 ($<0.001$) & -0.45 ($<0.001$) \\
\bottomrule
\end{tabular}
\end{table}

\subsection{Simulations on empirical networks}

\begin{figure}[H]
\centering
\begin{subfigure}[b]{0.47\textwidth}
\centering
\includegraphics[width=\textwidth]{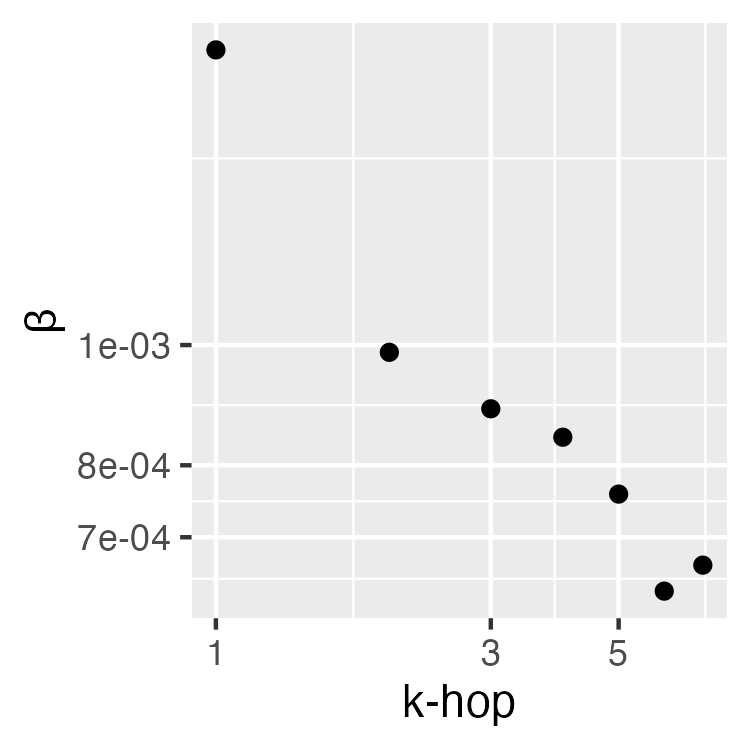}
\caption{E-mail}
\end{subfigure}
\hfill
\begin{subfigure}[b]{0.47\textwidth}
\centering
\includegraphics[width=\textwidth]{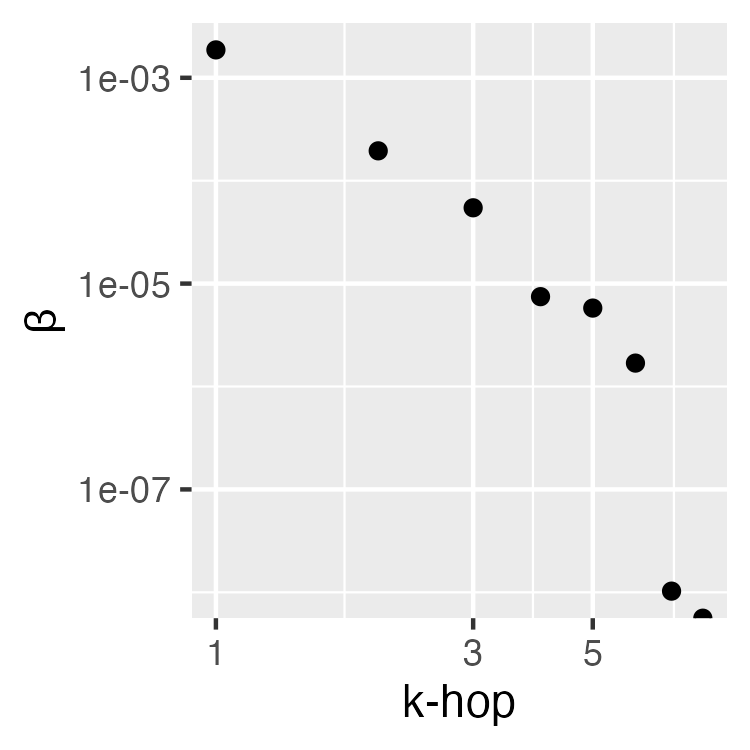}
\caption{Facebook}
\end{subfigure}

\begin{subfigure}[b]{0.47\textwidth}
\centering
\includegraphics[width=\textwidth]{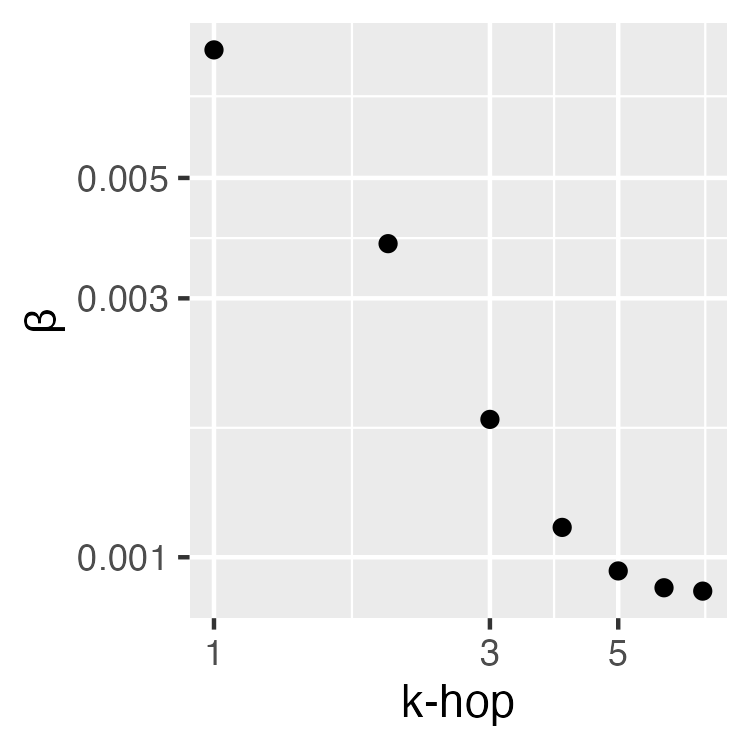}
\caption{Pigg Party}
\end{subfigure}
\hfill
\begin{subfigure}[b]{0.47\textwidth}
\centering
\includegraphics[width=\textwidth]{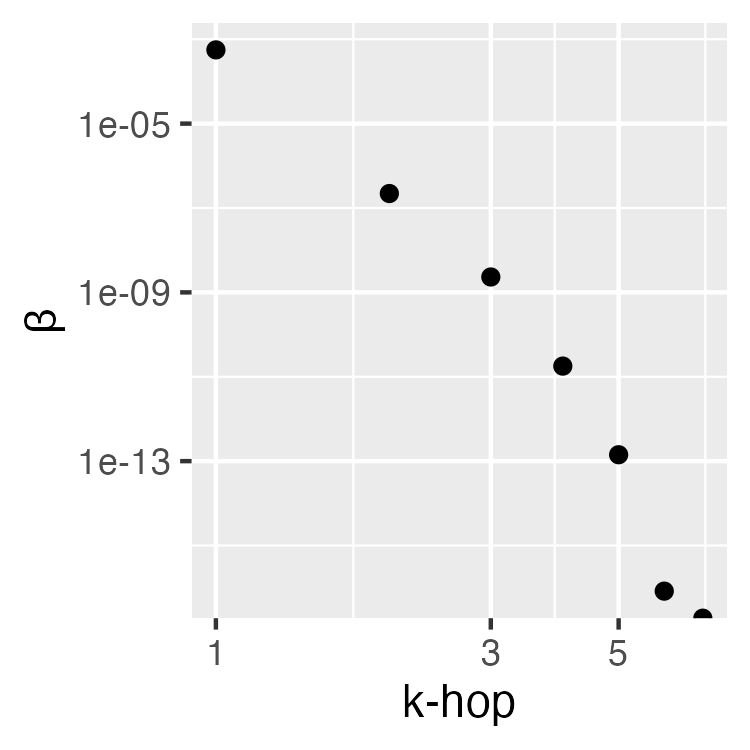}
\caption{Twitter}
\end{subfigure}
\caption{\textbf{Simulated distance-decay curves on empirical social networks.} Points show the simulated coefficient $\beta_k$ as a function of hop distance on log-log axes.}
\label{fig_realdata_sim}
\end{figure}

We also applied the same individual-based model to empirical social networks: Pigg Party, Twitter (now X), Facebook, and a university e-mail network. The resulting decay curves are shown in Fig.~\ref{fig_realdata_sim}, and the model comparison is summarized in Table~\ref{tbl_real_model_test}. For all networks, we used $k\leq6$ for the regression comparison.

In the sparse simulation approximation of Pigg Party and in the Twitter network, adding the fitted values from the power-decay model to the exponential model significantly improved the fit, whereas adding the fitted values from the exponential model to the power-decay model did not. The power-decay model was therefore relatively supported for these two networks. For the E-mail and Facebook networks, the exponential model was not rejected at the 5\% level, although the results showed weaker evidence in the same direction.

\begin{figure}[H]
\centering
\includegraphics[width=0.6\textwidth]{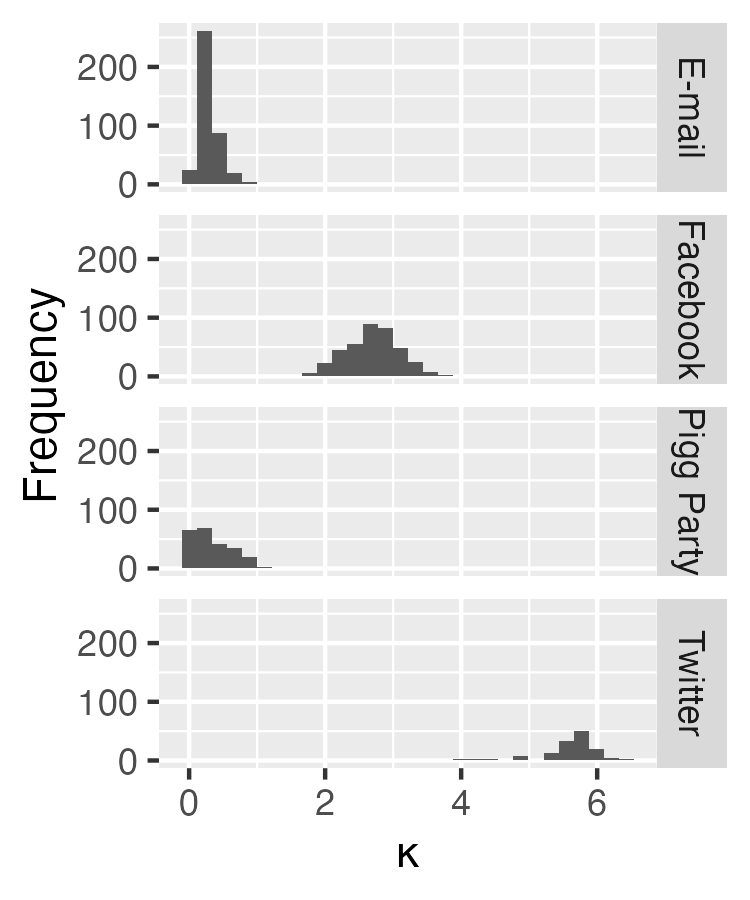}
\caption{\textbf{Distribution of source-specific decay rates on empirical social networks.} The distributions summarize $\kappa_s$ derived from source-level distance profiles using the exponential form in Eq.~\eqref{eq:source_decay}.}
\label{fig_kappa_freq_realdata}
\end{figure}

The relationship between $\kappa_s$ and standard centrality measures was less consistent in empirical networks than in the model-generated networks (Fig.~\ref{fig_kappa_freq_realdata} and Table~\ref{tbl_real_kappa_cor}). E-mail and Facebook showed patterns broadly similar to those of the BA and ER networks, except for the clustering coefficient on Facebook. In Pigg Party and Twitter, however, the correlations with centrality measures were weak or, in some cases, reversed. This suggests that empirical online social networks exhibit structural features not captured by simple generative models, and that these features can also shape observed distance-decay patterns.

\begin{table}[H]
\centering
\caption{\textbf{Descriptive comparison of simulated decay models on empirical networks.} The evaluation range was $k\leq6$ for all networks.}
\label{tbl_real_model_test}
\small
\begin{tabular}{llllrrr}
\toprule
Network & Null model tested & Adj. $R^2$ & Added fitted & Estimate & S.E. & $P$ value \\
\midrule
E-mail & $M_e$: $\log\beta_k\sim k$ & 0.703 & fitted($M_p$) & 1.413 & 0.503 & 0.067 \\
E-mail & $M_p$: $\log\beta_k\sim\log k$ & 0.898 & fitted($M_e$) & -0.454 & 0.534 & 0.458 \\
Facebook & $M_e$: $\log\beta_k\sim k$ & 0.947 & fitted($M_p$) & 0.694 & 0.254 & 0.071 \\
Facebook & $M_p$: $\log\beta_k\sim\log k$ & 0.977 & fitted($M_e$) & 0.321 & 0.257 & 0.300 \\
Pigg Party & $M_e$: $\log\beta_k\sim k$ & 0.868 & fitted($M_p$) & 1.218 & 0.256 & 0.018 \\
Pigg Party & $M_p$: $\log\beta_k\sim\log k$ & 0.981 & fitted($M_e$) & -0.237 & 0.269 & 0.443 \\
Twitter & $M_e$: $\log\beta_k\sim k$ & 0.897 & fitted($M_p$) & 1.067 & 0.240 & 0.021 \\
Twitter & $M_p$: $\log\beta_k\sim\log k$ & 0.986 & fitted($M_e$) & -0.072 & 0.249 & 0.791 \\
\bottomrule
\end{tabular}
\end{table}

\begin{table}[H]
\centering
\caption{\textbf{Correlations between $\kappa_s$ and node-level network statistics in empirical networks.} Values in parentheses are $P$ values.}
\label{tbl_real_kappa_cor}
\small
\begin{tabular}{lrrrr}
\toprule
Indicator & E-mail & Facebook & Pigg Party & Twitter \\
\midrule
Degree & -0.23 (0.000) & -0.54 (0.000) & 0.19 (0.004) & 0.01 (0.945) \\
Clustering & -0.03 (0.528) & 0.26 (0.000) & 0.16 (0.012) & -0.25 (0.003) \\
Betweenness & -0.13 (0.009) & -0.12 (0.019) & -0.08 (0.203) & -0.01 (0.934) \\
Closeness & -0.32 (0.000) & -0.59 (0.000) & 0.06 (0.338) & -0.21 (0.013) \\
PageRank & -0.17 (0.001) & -0.46 (0.000) & 0.06 (0.338) & 0.00 (0.969) \\
\bottomrule
\end{tabular}
\end{table}

\section{Discussion}

This study combined behavioral log-based inference, empirical network scaling analysis, and individual-based modeling to characterize distance-dependent associations in perceived online social support within an avatar communication network. We showed that perceived online social support could be inferred from behavioral and profile signals with moderate predictive accuracy. The predictive component should be understood as a measurement-enabling step rather than as a stand-alone model for individual assessment: its test-set performance ($r=0.362$, $R^2=0.122$) indicates a moderate but incomplete signal. If prediction errors were approximately classical and independent, they would generally attenuate distance-specific associations rather than create them. We therefore treated the inferred scores as noisy proxies and restricted our claims to aggregate distance-dependent patterns. We then found that inferred support scores were positively associated across multiple hops in the Pigg Party interaction network, even after adjusting for the target user's baseline score and covariates. Over the observed range, the adjusted association decayed more slowly than a single exponential and was better described by a power-decay model. These findings indicate that support-related associations in an online community can extend beyond immediately adjacent users.

A conceptual contribution of the study is to treat perceived online social support as a measurable psychological state on a network. Previous network and statistical physics models have often modeled social support as a resource that alters recovery probabilities or state-transition rates in epidemic dynamics~\cite{Chen2018SS,Zhou2019}.
Our empirical analysis instead starts from questionnaire-based measurements of perceived support and uses platform traces to estimate each active user's perceived online social support score. We then ask how these estimated scores are associated with later perceived support at different graph distances.
This perspective bridges survey-based online support research, digital trace-based psychological inference, and network-scaling analysis.

The observed distance-decay pattern suggests that perceived online social support is not confined to immediately adjacent dyads but is organized across broader regions of the interaction network. One plausible interpretation is that support-related processes are chained through social ties. A user who receives supportive interactions may become more likely to provide support to another user, a mechanism related to pay-it-forward behavior and upstream reciprocity~\cite{Nowak2006b,Bartlett2006,Iwagami2010,Horita2016,Obayashi2023}. This interpretation is also consistent with broader evidence on emotional-support and cooperative-behavior cascades in social networks~\cite{Lakon2017,Fowler2010}. The present study adds a distance-decay description to this literature by estimating how the magnitude of a longitudinal
support association changes from one hop to several hops in a large online interaction network. However, this interpretation should remain descriptive. The analysis did not observe explicit sequences of support provision.

The modeling results provide one possible mechanism for the heavy-tailed pattern. A mixture of source-specific exponential decays with heterogeneous decay rates can produce an aggregate curve that decays more slowly than a single exponential~\cite{Beck2003,Mitzenmacher2003}. In the individual-based simulations, the same local updating rule generated different distance-decay patterns depending on network structure. BA networks and some empirical networks produced curves closer to a power-decay form, whereas a single exponential better described the WS network. These findings suggest that the distance-decay profile of a psychosocial association depends not only on the local updating rule but also on heterogeneity in network position and reachability.

The new framing also clarifies what the individual-based model does and does not explain. The model shows that heavy-tailed aggregate decay can arise even when the local source-specific profiles are approximated by exponential decay. This is compatible with the broader observation that social communication networks have heterogeneous tie strengths, activity levels, and weighted paths~\cite{Barrat2004,Onnela2007,Iniguez2023,Heydari2024}. It is also compatible with work showing that burstiness and edge weights can slow spreading processes~\cite{Karsai2011,Iribarren2009,Min2011}. Nevertheless, our simulations used simplified unweighted networks and an event-driven state update that did not encode message content, relationship quality, temporal burstiness, or heterogeneous activity schedules. The model should therefore be understood as a minimal explanation based on source-level decay-rate heterogeneity, not as a complete behavioral model of online support exchange.

The study has some limitations. First, the longitudinal regression does not establish causality. Quasi-experimental designs or platform interventions would be needed to test causal propagation.
Second, the empirical analysis relies on inferred support scores, and the prediction accuracy was moderate rather than high.
Third, the individual-based model deliberately simplifies behavior, content, and relationship strength. It does not explicitly include weighted edges, temporal changes in the network, homophily, or selective exposure. In addition, support provision can itself increase perceived support~\cite{Tyler2006}, and the effect of received support on perceived support can depend on the recipient's state~\cite{Melrose2015}; these mechanisms are not yet included in the model.

Future research should connect the empirical and modeling components more directly by estimating source-specific decay rates in the empirical data, relating these rates to network position, and extending the model to weighted and temporal networks with heterogeneous activity. Replication on other online platforms will also be necessary to determine whether the observed distance-decay pattern is specific to avatar communication or reflects a broader property of psychosocial states in online social networks.

\section{Methods}

\subsection{Avatar communication application}

Pigg Party is a Japanese avatar communication application and virtual community operated by CyberAgent, Inc. Players communicate synchronously through personalized avatars in virtual spaces (Fig.~\ref{fig_piggparty}). In addition to text messages, they can use avatar actions that express emotions and gestures. Each player has a private room and can visit common spaces, invite other users to their own room, or visit other users' rooms. Pigg Party uses a stylized, non-realistic two-dimensional avatar style. Previous work reported that 61\% of players were teenagers and 65\% were female~\cite{MasanoriTakano2019,takano2023_complexity}.

\subsection{Participants and survey waves}

We conducted two online surveys of Pigg Party players in 2020. Wave 1 was conducted from 26 April to 2 May, and Wave 2 was conducted from 1 to 7 June. Participants provided informed consent for academic use of their responses. Recruitment was conducted through the application. In Wave 1, an in-app announcement was sent to all players, and respondents received virtual coins equivalent to 100 JPY. In Wave 2, Wave 1 participants were invited to a follow-up survey with the same incentive. Participants entered their Pigg Party ID in the questionnaire, thereby linking their survey responses to behavioral logs and social network data.

We excluded the fastest 3\% of complete respondents, who were considered likely to include inattentive responses (completion time $<4.10$ min). We also excluded users whose total application usage in the preceding month was below the 2.5th percentile of the sample, as their behavioral data were deemed insufficient. After applying these criteria, the analytical sample included $N=2,923$ users in Wave 1 and $N=745$ users in Wave 2.

\subsection{Perceived online social support}

We measured perceived emotional and instrumental support from online friends on Pigg Party. The measure used established social-support scales with confirmed reliability and validity~\cite{Fukuoka1997}. Each source included 12 items: six emotional-support items and six instrumental-support items. Example items were ``Cheer me up when I am depressed'' for emotional support and ``When I suddenly need a few thousand yen because I lost my wallet or must pay for something that has been damaged, lend me the money'' for instrumental support.

In this study, perceived online social support refers to perceived emotional and instrumental support from online friends on Pigg Party. Responses were measured on a five-point Likert scale ranging from 1, ``Pigg Party friends would never do it,'' to 5, ``Pigg Party friends would likely do it.'' We performed confirmatory factor analysis using maximum likelihood estimation. Internal consistency was high: Cronbach's $\alpha$ was 0.947 and 0.952 for emotional support, and 0.964 and 0.966 for instrumental support, in Wave 1 and Wave 2, respectively. Because emotional and instrumental support were highly correlated and following previous work~\cite{Semmer2008,Shakespeare-Finch2011,takano_icwsm2022,takano_icwsm2025}, we performed a principal component analysis of online social support. We used the first principal component as the overall support-strength score.

\subsection{Behavioral and profile features}
Features were constructed from behavioral logs during the four weeks immediately preceding each wave-specific reference date: 26 April 2020 for Wave 1 and 31 May 2020 for Wave 2. The feature set included counts of in-app actions during the corresponding one-month period; dominant usage time during the corresponding one-month period; avatar age and gender, which do not necessarily reflect the user's real attributes; total usage time during the corresponding one-month period; ten-dimensional Paragraph2Vec embeddings of avatar clothing and accessories; ten-dimensional Paragraph2Vec embeddings of avatar facial parts; the total number of owned avatar clothing and accessory items; room size and number of interior items; the number of mutual follows; and survey wave.

The feature set also included relationship and network-based features. Degree and mean edge weight were computed from a visit-based interaction network constructed from private-room visit logs during the same one-month period. In this network, nodes represented users; an edge connected a visitor to the owner of the visited room; and the edge weight was defined as the aggregated dwell time in seconds. Mean edge weight, therefore, represented the average strength of a user's room-visit interactions.

Mutual follows, degree, and mean edge weight represent the local interaction environment. These variables are theoretically relevant because perceived online social support is expected to be grounded in the frequency of local contact and relationship strength. However, because later network-distance analyses used a network constructed from the same interaction logs, these features could introduce local network autocorrelation into the inferred scores.

\subsection{Prediction model selection}

We compared random forest, LightGBM, Elastic Net regression, and a linear-kernel support vector machine. The user partitioned the 3,668 wave-level observations into training (80\%), validation (10\%), and test (10\%) sets, so observations from the same user could not appear in more than one partition. For each algorithm, hyperparameters were tuned by grid search using the training data, and the configuration with the lowest validation-set mean absolute error was retained. The four tuned algorithms were then compared on the user-separated test set using mean absolute error, root mean squared error, $R^2$, and the correlation between predicted and observed scores. Although LightGBM achieved the lowest mean absolute error, the random forest achieved the highest correlation, the lowest root mean squared error, and the highest $R^2$. Because the downstream network analysis depended on retaining between-user covariation in perceived support rather than minimizing absolute error alone, we selected the random forest as the final inference model. The test set was used for this internal comparison of candidate algorithms; the reported metrics should therefore not be interpreted as an independent external validation. The final random forest was then used to infer perceived online social support scores for active nonrespondents in each wave.

\subsection{Social network construction}

We constructed online social networks from logs of visits to other users' private rooms, following previous work~\cite{takano2023_complexity}. We assumed an edge between a visitor and the room owner because visitors tend to be friends of the owner~\cite{MasanoriTakano2019}. This visit-based network approximates social contact rather than directly observing message paths, and has been used in previous analyses of Pigg Party interaction patterns~\cite{takano2023_complexity}. Edge weights were defined by aggregated dwell time in seconds. For network-based features used in the prediction model, we constructed wave-specific networks using logs from the four weeks immediately preceding each wave-specific reference date: 29 March to 25 April 2020 for Wave 1 and 3 May to 31 May 2020 for Wave 2. In contrast, the network-distance association analysis used only the Wave 1 interaction network.

The Wave 1 network used in the network-distance association analysis had a diameter of 24. Its degree distribution had a power-law exponent of 4.80, which is larger than those of many social networks~\cite{Barabasi2016}. These properties indicate that the network was relatively similar to a random network, with fewer prominent hub nodes than in typical scale-free social networks. In this network, the average fraction of users reachable within $k$ hops from a node was 45.4\% at five hops and 70.3\% at six hops.

\subsection{Network-distance association analysis}

For each hop distance $k$, we examined whether the Wave 1 support score of a source user $j$, denoted by $y_j^{(1)}$, was predictively associated with the Wave 2 support score of a target user $i$, denoted by $y_i^{(2)}$. We controlled for the target user's Wave 1 score $y_i^{(1)}$ and covariates:
\begin{equation}
y_i^{(2)} = \alpha y_i^{(1)} + \beta_k y_j^{(1)} + \boldsymbol{\zeta}^{\top}\mathbf{C}_i + \varepsilon,\quad d(i,j)=k.
\label{eq:diffusion}
\end{equation}
The covariate vector $\mathbf{C}_i$ included the target user's avatar age, avatar gender, degree, mean edge weight, number of owned clothing and accessory items, room size, number of interior items, and recent usage time. The coefficient $\beta_k$ represents the adjusted longitudinal association between a source user's Wave 1 support score at hop distance $k$ and the target user's Wave 2 support score. The distance $d(i,j)$ was defined as the shortest-path length in the Wave 1 interaction network.

Because the number of pairs $(i,j)$ increases rapidly with hop distance, we estimated $\beta_k$ by repeated sampling rather than using all possible pairs. In each repetition, we randomly sampled 2,000 source nodes $j$ and estimated Eq.~\eqref{eq:diffusion} using pairs satisfying $d(i,j)=k$. We repeated this procedure 100 times and used the mean of $\beta_k$ across repetitions as the final estimate. Since the same target or source can appear in multiple pairs, standard errors based on pairwise independence may be underestimated. We therefore focused on the average coefficients and the distance-decay shape rather than on confidence intervals for individual coefficients.

To compare the distance-decay pattern, we fitted two linear models: an exponential-decay model with dependent variable $\log\beta_k$ and predictor $k$, and a power-decay model with dependent variable $\log\beta_k$ and predictor $\log k$. Because $\beta_k$ can be affected by network size and the number of reachable nodes, we visually inspected the decay curve and used a range in which this effect was considered small. We compared the non-nested models using the Davidson-MacKinnon $J$ test, after checking the coefficient estimates and goodness of fit.

\subsection{Individual-based model}

We represented interpersonal relations as an undirected graph $G=(V,E)$ with $|V|=n$. Each node $i\in V$ has a perceived-support state $y_i(t)\in\mathbb{R}_{\geq0}$. Time is discrete, $t=1,\ldots,T$, and one message event, sender to receiver, occurs at each step.
Receiving a message is interpreted as perceiving it as a support resource available when needed.
The initial state $y_i(0)$ corresponds to Wave 1, and the final state $y_i(T)$ corresponds to Wave 2.

\subsubsection{Messaging process}

A random walk with reply and restart probabilities generated the message sequence. Let $i_t$ be the sender at time $t$ and $p_t$ be the sender at the previous step, that is, the previous interaction partner. At each step, the conversation thread restarts with probability $r\in[0,1)$; then $i_t$ is chosen uniformly at random and $p_t=\emptyset$. If the thread does not restart and $p_t\in N(i_t)$, the sender chooses the previous partner as the receiver, $j_t=p_t$, with fixed reply probability
\begin{equation}
\Pr(\mathrm{reply})=p_r.
\end{equation}
If the sender does not reply, the receiver $j_t$ is chosen uniformly from $N(i_t)\setminus\{p_t\}$; if this set is empty, the receiver is chosen uniformly from $N(i_t)$. The process then updates $(p_{t+1},i_{t+1})\leftarrow(i_t,j_t)$. This random-walk representation of interaction chains is consistent with formulations of upstream reciprocity as chains of prosocial actions on graphs~\cite{Nowak2006b}.

\subsubsection{State update}

When the message $i_t\to j_t$ occurs, only the receiver is updated:
\begin{equation}
y_{j_t}(t+1)=(1-\mu)y_{j_t}(t)+\Delta y_{i_t}(t),
\label{eq:update}
\end{equation}
where $0\leq\mu<1$ and $\Delta>0$. All other nodes satisfy $y_\ell(t+1)=y_\ell(t)$. The parameter $\mu$ is interpreted not as a natural decay that acts on all nodes at each time but as the degree to which the receiver's existing state is diluted by new input during a receiving event. The model is therefore an approximation of event-driven state updating rather than a model of temporal forgetting.

\subsubsection{Distance-specific effect size}

Let $d(s,j)$ be the shortest-path distance between source $s$ and node $j$. We defined the sensitivity to a small intervention $\epsilon>0$ added to the initial value of the source $s$ as
\begin{equation}
\delta_{j\mid s}=\frac{y^{(\mathrm{int})}_j(T)-y^{(\mathrm{base})}_j(T)}{\epsilon},
\end{equation}
using the same network and the same message sequence for the intervention and baseline trajectories. The source-specific effect averaged over the distance-$k$ shell was
\begin{equation}
\beta_{k\mid s}=\frac{1}{|\{j:d(s,j)=k\}|}\sum_{j:d(s,j)=k}\delta_{j\mid s}.
\end{equation}
For a set of sources $S$, we aggregated source-specific effects as $\beta_k=|S|^{-1}\sum_{s\in S}\beta_{k\mid s}$, or equivalently by pair-weighted averaging where appropriate. We compared exponential and power-decay models using the same log-linear regressions and $J$ test as in the empirical distance-decay analysis.

\subsection{Networks used for simulations}

For model-generated networks, we used BA scale-free networks~\cite{Barabasi1999}, ER random networks~\cite{Erdos1959} and WS small-world networks~\cite{Watts1998}. The number of nodes was $N=5000$, and parameters were chosen so that the average degree was approximately 6. The BA model used $m=3$ edges for each added node, the ER model used edge probability $p_{\mathrm{ER}}=0.0012$, and the WS model used rewiring probability $p_{\mathrm{WS}}=0.01$ and neighborhood size $k_{\mathrm{WS}}=6$.

For empirical networks, we used the largest connected components of the Pigg Party (sparse simulation approximation), Twitter Higgs, Facebook CMU, and university email networks. The external Twitter, Facebook, and email networks were obtained from the Network Repository~\cite{nr-aaai15}. The Pigg Party network used the Wave 1 room-visit data described above. Because the simulation model was designed for broad applicability and did not use edge weights, we constructed an unweighted approximation of the Pigg Party network by sampling each edge with probability proportional to its weight, using a proportionality coefficient of 0.01. This procedure preferentially preserves long dwell-time edges while retaining some shorter interactions. The resulting Pigg Party network is not identical to the weighted network used in the empirical analysis; it is a sparse approximation intended for mechanistic exploration.

\begin{table}[H]
\centering
\caption{\textbf{Summary statistics of empirical social networks used in simulations.}}
\label{tbl_smr_net}
\small
\begin{tabular}{lrrrr}
\toprule
Network & Pigg Party & Twitter & Facebook & E-mail \\
\midrule
Number of nodes & 1,109 & 456,293 & 6,621 & 1,133 \\
Number of edges & 1,179 & 12,508,249 & 249,959 & 5,451 \\
Mean degree & 2.126 & 54.826 & 75.505 & 9.622 \\
Clustering coefficient & 0.079 & 0.199 & 0.287 & 0.254 \\
Mean shortest-path length & 26.466 & 3.183 & 2.738 & 3.606 \\
\bottomrule
\end{tabular}
\end{table}

\subsection{Simulation parameters}

Unless otherwise noted, simulations used $p_r=0.167$, $r=0.02$, $\mu=0.05$, $\Delta=0.05$ and 500,000 steps. These values were chosen as a simple baseline for examining how the shape of distance decay depends on network structure. Sensitivity analyses varied the reply probability, restart probability, update weight, transmission strength, number of steps, source-set size, and random seed (Supplementary Note~2). BA networks showed stronger support for the power-decay model across many conditions, whereas ER networks showed weaker evidence in the same direction. In WS networks, the exponential model often had a higher adjusted $R^2$ than the power-decay model, although the $J$ test indicated that neither simple model was adequate under some conditions. Overall, results were relatively stable with respect to $p_r$, $r$, source-set size, and random seed, but depended on update weight, transmission strength, and observation time scale.

\section*{Ethics approval}

The survey and log-linkage analysis were approved by the Ethics Committee of the Graduate School of Technology, Industrial and Social Sciences, Tokushima University (ethics review reference no.~210). All procedures were conducted in accordance with guidelines for studies involving human participants, the ethical standards of the institutional research committee, and the 1964 Declaration of Helsinki and its later amendments.

Participants provided informed consent to participate in the survey and could stop at any time. They could also withdraw their responses after completing the survey. Participants were informed that the survey was conducted for academic research on mental health and bullying victimization. The informed consent form included a point of contact for requests to disclose or withdraw responses.

The Pigg Party application provider collected log and questionnaire data in accordance with its terms of service, privacy policy, and the survey-informed consent. The provider explicitly stated the purpose and scope of the collection and use of log and survey data in the informed consent form. The provider explained that the data would be used solely for academic research and not for business purposes. The authors who received the data were explicitly identified to participants in the informed-consent form. The provider supplied participant log data to the authors after removing identifying information.

All Pigg Party players, including those who did not participate in the survey, accepted the terms of service and privacy policy, which permitted the analysis of behavioral data for service improvement and academic research. The data were pseudonymized, and identifying information was removed. Quantitative outputs are reported only in aggregate form, and no identifying information is presented.

\section*{Data availability}

The Pigg Party survey and behavioral-log data are not publicly available because they contain sensitive user-level behavioral information and were provided under informed consent and data-use restrictions. The external network datasets used in the simulations are publicly available from the Network Repository.

\section*{Code availability}

The simulation code (R script) is provided as S1 Code.

\section*{Acknowledgements}

This work was supported by JST PRESTO Grant Number JPMJPR2367. The authors thank Prof. Eizo Akiyama of the University of Tsukuba for valuable comments on the interpretation of the study.

\section*{Author contributions}
M.T. conceived the study, conducted the analyses, interpreted the results, and wrote the manuscript. K.Y. developed the questionnaire used in the survey and empirical analysis and contributed to the interpretation of the empirical results. M.C. reviewed the mathematical and simulation analyses and contributed to their interpretation. F.T. contributed to the interpretation of the results. All authors reviewed and approved the manuscript.

\section*{Competing interests}

M.T. is employed by CyberAgent, Inc., the provider of Pigg Party.
K.Y. and F.T. were funded by CyberAgent, Inc.

\end{document}


\section*{Supplementary Note 1. Sensitivity analysis excluding adjacent dyads in the empirical analysis}
\label{supp:study1_k_sensitivity}

The main empirical analysis used inferred perceived online social support scores predicted from behavioral and profile features, including local interaction and network-related variables such as mutual-follow counts, degree, and mean edge weight.
Including these variables is theoretically justified because perceived online social support in an avatar-based communication service is grounded in local social interactions and repeated contact.
At the same time, because the same interaction records are also used to construct the social network, the coefficient at $k=1$ could be especially sensitive to local network similarity or to the use of local network features in the prediction stage.

The selected random forest achieved a correlation of $r=0.362$ and $R^2=0.122$ on the user-separated test set used for internal model comparison. These values indicate that the inferred scores captured a moderate but incomplete signal and should be interpreted as noisy proxies rather than replacements for questionnaire scores. If prediction errors were classical and independent across users, they would be expected mainly to attenuate cross-user associations. That assumption is not guaranteed here: predictors such as mutual follows, degree, and mean edge weight share an interaction-log data source with the network used to define distance, so prediction errors may be spatially structured. The following analysis therefore addresses the targeted question of whether the decay-model comparison is driven solely by the adjacent-user coefficient; it does not establish that the inferred scores are free of network autocorrelation.

To examine whether the main conclusion was driven only by immediately adjacent users, we repeated the descriptive decay-model comparison after excluding $k=1$ and fitting the models only over $2\le k\le6$.
This analysis should be interpreted as a sensitivity check on the distance range rather than as a full removal of network autocorrelation in the predicted scores.
For consistency with the main text, we compared the exponential model $M_e:\log\beta_k\sim k$ with the power-law-like model $M_p:\log\beta_k\sim\log k$ using adjusted $R^2$ and the Davidson--MacKinnon J test.

The results are shown in Table~\ref{tbl_pg_model_test_k2_6}.
The power-law-like model had a higher adjusted $R^2$ than the exponential model ($0.973$ vs. $0.905$).
In the J test, adding the fitted values from $M_p$ to $M_e$ provided marginal additional explanatory power, whereas adding the fitted values from $M_e$ to $M_p$ did not improve the model.
Thus, the slower-than-exponential pattern was not solely attributable to the $k=1$ adjacent-user coefficient.
Because the analysis uses only five distance points, however, this result should be regarded as descriptive evidence for a heavy-tailed or power-law-like decay over the observed range, not as a definitive statistical test of a power law.

\begin{table}[H]
\centering
\caption{Descriptive comparison of decay models over $2\le k\le6$ in the empirical analysis. $M_e$: exponential model, $\log \beta_k \sim k$. $M_p$: power-law-like model, $\log \beta_k \sim \log k$.}
\label{tbl_pg_model_test_k2_6}
\scriptsize
\setlength{\tabcolsep}{3pt}
\begin{tabular}{ l | l| rrrr|r}
\toprule
Null model tested & Adj. $R^2$ & Added fitted & Estimate & S.E. & p-value & Interpretation\\ \midrule
$M_e$: $\log \beta_k \sim k$ & 0.905& fitted($M_p$)& 1.528 & 0.512& 0.096 & Marginal evidence against $M_e$\\
$M_p$: $\log \beta_k \sim \log k$ & 0.973 &  fitted($M_e$)& -0.549 &   0.526 &  0.406& Do not reject $M_p$\\ 
\bottomrule
\end{tabular}
\end{table}

\section*{Supplementary Note 2. Parameter sensitivity analysis for the individual-based model}
\label{supp:study2_parameter_sensitivity}

To examine whether the individual-based model depended on a narrow parameter setting, we conducted one-at-a-time sensitivity analyses around the baseline
\[
\begin{aligned}
p_r&=0.167, & r&=0.02, & \mu&=0.05,\\
\Delta&=0.05, & T&=500{,}000, & |S|&=400,
\end{aligned}
\]
with random seed 423.
Unless otherwise varied, the source-set size was fixed at 400 and the random seed was fixed at 423.
The random seed controlled network generation, source-node sampling, and message-sequence generation.
We varied the reply probability $p_r\in\{0,0.1,0.167,0.3\}$ and restart probability $r\in\{0.005,0.02,0.05\}$. We also varied the update weight $\mu\in\{0.01,0.05,0.1\}$ and transmission strength $\Delta\in\{0.01,0.05,0.1\}$. The remaining variations were the number of steps $T\in\{100{,}000,500{,}000,1{,}000{,}000\}$, source-set size $\{200,400,800\}$, and random seed $\{423,846,1269\}$.
The 22 parameter settings are one-at-a-time evaluations around the baseline; because the baseline value appears within multiple parameter families, the counts should be interpreted as descriptive sensitivity-analysis units rather than independent replications.

For each condition, we estimated $\beta_k$ on BA, ER, and WS networks and compared an exponential model, $M_e:\log\beta_k\sim k$, with a power-law-like model, $M_p:\log\beta_k\sim\log k$.
The fitted range was the same as in the main text: $k\le6$ for BA and ER networks, and $2\le k\le14$ for WS networks.

Table~\ref{tbl_sens_overall} summarizes the results.
``$M_e$ rejected by J test'' and ``$M_p$ rejected by J test'' indicate the number of conditions in which the Davidson--MacKinnon J test rejected the exponential or power-law-like model at the 5\% level.
Because the J test evaluates whether the fitted values from one non-nested model add explanatory power to the other, both models can be rejected in some parameter settings.
The rejection counts should therefore be interpreted descriptively and together with adjusted $R^2$, not as definitive model-selection probabilities.

In BA networks, the power-law-like model had higher adjusted $R^2$ than the exponential model in 18 of 22 conditions, with a median difference in adjusted $R^2$ of 0.074.
In ER networks, the same direction appeared in 20 of 22 conditions, but the median difference was small (0.007), indicating a weaker and more marginal pattern.
In WS networks, the exponential model generally provided a better descriptive fit than the power-law-like model in terms of adjusted $R^2$; however, the J-test results indicate that neither simple functional form was uniformly adequate across all parameter settings.
Thus, the heavy-tailed distance decay in BA networks was relatively stable to $p_r$, $r$, source-set size, and random seed, while some settings of $\mu$, $\Delta$, and $T$ changed the relative fit.

\begin{table}[H]
\centering
\caption{Overall summary of the parameter sensitivity analysis for the individual-based model. $\Delta R^2_{\mathrm{adj}}$ is the adjusted $R^2$ of the power-law-like model minus that of the exponential model. $\mathrm{CV}(\kappa_s)$ is the coefficient of variation of source-specific decay rates. $M_e$: exponential model, $\log \beta_k \sim k$. $M_p$: power-law-like model, $\log \beta_k \sim \log k$.}
\label{tbl_sens_overall}
\scriptsize
\setlength{\tabcolsep}{3pt}
\resizebox{\textwidth}{!}{%
\begin{tabular}{lrrrrrr}
\toprule
Network & Conditions & Power-law-like better & Median $\Delta R^2_{\mathrm{adj}}$ [range] & $M_e$ rejected by J test & $M_p$ rejected by J test & Median $\mathrm{CV}(\kappa_s)$ [range] \\
\midrule
BA & 22 & 18/22 & 0.074 [-0.149, 0.164] & 21/22 & 5/22 & 0.124 [0.056, 0.214] \\
ER & 22 & 20/22 & 0.007 [-0.046, 0.156] & 1/22 & 1/22 & 0.145 [0.107, 0.304] \\
WS & 22 & 1/22 & -0.091 [-0.096, 0.099] & 18/22 & 21/22 & 0.043 [0.041, 0.068] \\
\bottomrule
\end{tabular}
}
\end{table}

Table~\ref{tbl_sens_by_parameter} reports summaries by parameter family.
For BA networks, the power-law-like model had higher adjusted $R^2$ in all conditions varying $p_r$, $r$, random seed, and source-set size.
The same four parameter families showed the same direction in ER networks, although the differences were smaller.
For WS networks, the exponential model usually had higher adjusted $R^2$ than the power-law-like model, and $\mathrm{CV}(\kappa_s)$ was smaller than in BA and ER networks.
These results are consistent with the interpretation that source-level heterogeneity in decay rates is one contributing factor to heavy-tailed aggregate decay.
However, the coefficient of variation of $\kappa_s$ alone does not fully determine the model comparison results; the shape of the $\kappa_s$ distribution, finite-size effects, reachability, and the association between $A_s$ and $\kappa_s$ may also matter.

\begin{table}[H]
\centering
\caption{Sensitivity-analysis summary by varied parameter family. Each row summarizes the conditions generated by varying one parameter family while holding the other parameters at the baseline.}
\label{tbl_sens_by_parameter}
\scriptsize
\setlength{\tabcolsep}{4pt}
\begin{tabular}{llrrr}
\toprule
Network & Varied parameter & Power-law-like better & Median $\Delta R^2_{\mathrm{adj}}$ [range] & Median $\mathrm{CV}(\kappa_s)$ [range] \\
\midrule
BA & $\Delta$ & 1/3 & -0.009 [-0.149, 0.074] & 0.124 [0.056, 0.158] \\
BA & $T$ & 2/3 & 0.074 [-0.016, 0.164] & 0.124 [0.067, 0.214] \\
BA & $\mu$ & 2/3 & 0.072 [-0.145, 0.074] & 0.120 [0.096, 0.124] \\
BA & $p_r$ & 4/4 & 0.075 [0.072, 0.077] & 0.124 [0.124, 0.128] \\
BA & $r$ & 3/3 & 0.073 [0.073, 0.074] & 0.124 [0.123, 0.126] \\
BA & seed & 3/3 & 0.074 [0.069, 0.084] & 0.130 [0.124, 0.138] \\
BA & source-set size & 3/3 & 0.074 [0.071, 0.074] & 0.124 [0.121, 0.135] \\
\midrule
ER & $\Delta$ & 2/3 & 0.007 [-0.019, 0.156] & 0.180 [0.144, 0.215] \\
ER & $T$ & 3/3 & 0.042 [0.007, 0.108] & 0.170 [0.144, 0.304] \\
ER & $\mu$ & 2/3 & 0.007 [-0.046, 0.121] & 0.144 [0.107, 0.200] \\
ER & $p_r$ & 4/4 & 0.008 [0.001, 0.015] & 0.153 [0.135, 0.184] \\
ER & $r$ & 3/3 & 0.007 [0.007, 0.007] & 0.144 [0.144, 0.155] \\
ER & seed & 3/3 & 0.007 [0.000, 0.007] & 0.145 [0.135, 0.145] \\
ER & source-set size & 3/3 & 0.007 [0.006, 0.009] & 0.149 [0.144, 0.149] \\
\midrule
WS & $\Delta$ & 1/3 & -0.092 [-0.095, 0.099] & 0.062 [0.043, 0.068] \\
WS & $T$ & 0/3 & -0.092 [-0.092, 0.000] & 0.043 [0.043, 0.047] \\
WS & $\mu$ & 0/3 & -0.092 [-0.096, -0.080] & 0.047 [0.043, 0.050] \\
WS & $p_r$ & 0/4 & -0.088 [-0.092, -0.080] & 0.044 [0.043, 0.045] \\
WS & $r$ & 0/3 & -0.089 [-0.092, -0.077] & 0.043 [0.042, 0.043] \\
WS & seed & 0/3 & -0.087 [-0.092, -0.081] & 0.043 [0.041, 0.044] \\
WS & source-set size & 0/3 & -0.090 [-0.092, -0.085] & 0.042 [0.042, 0.043] \\
\bottomrule
\end{tabular}
\end{table}

\begin{figure}[H]
\centering
\includegraphics[width=0.9\linewidth]{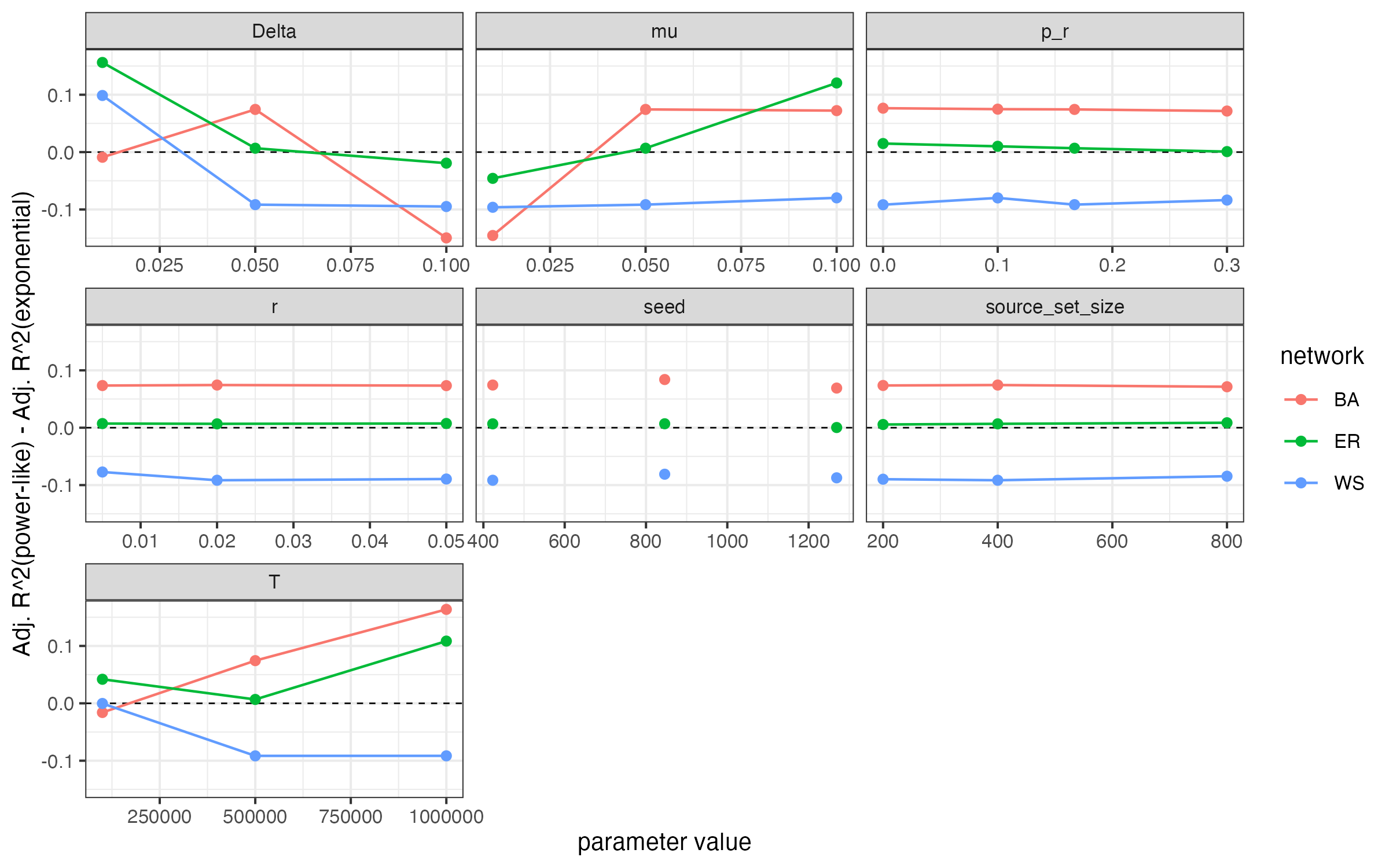}
\caption{Difference in adjusted $R^2$ between the power-law-like and exponential models for each parameter setting. Positive values indicate better descriptive fit of the power-law-like model. The $T$ panel is ordered numerically. In the seed panel, points are shown for comparison only; no ordered trend is implied.}
\label{fig_sens_fit_advantage}
\end{figure}

\begin{figure}[H]
\centering
\includegraphics[width=0.9\linewidth]{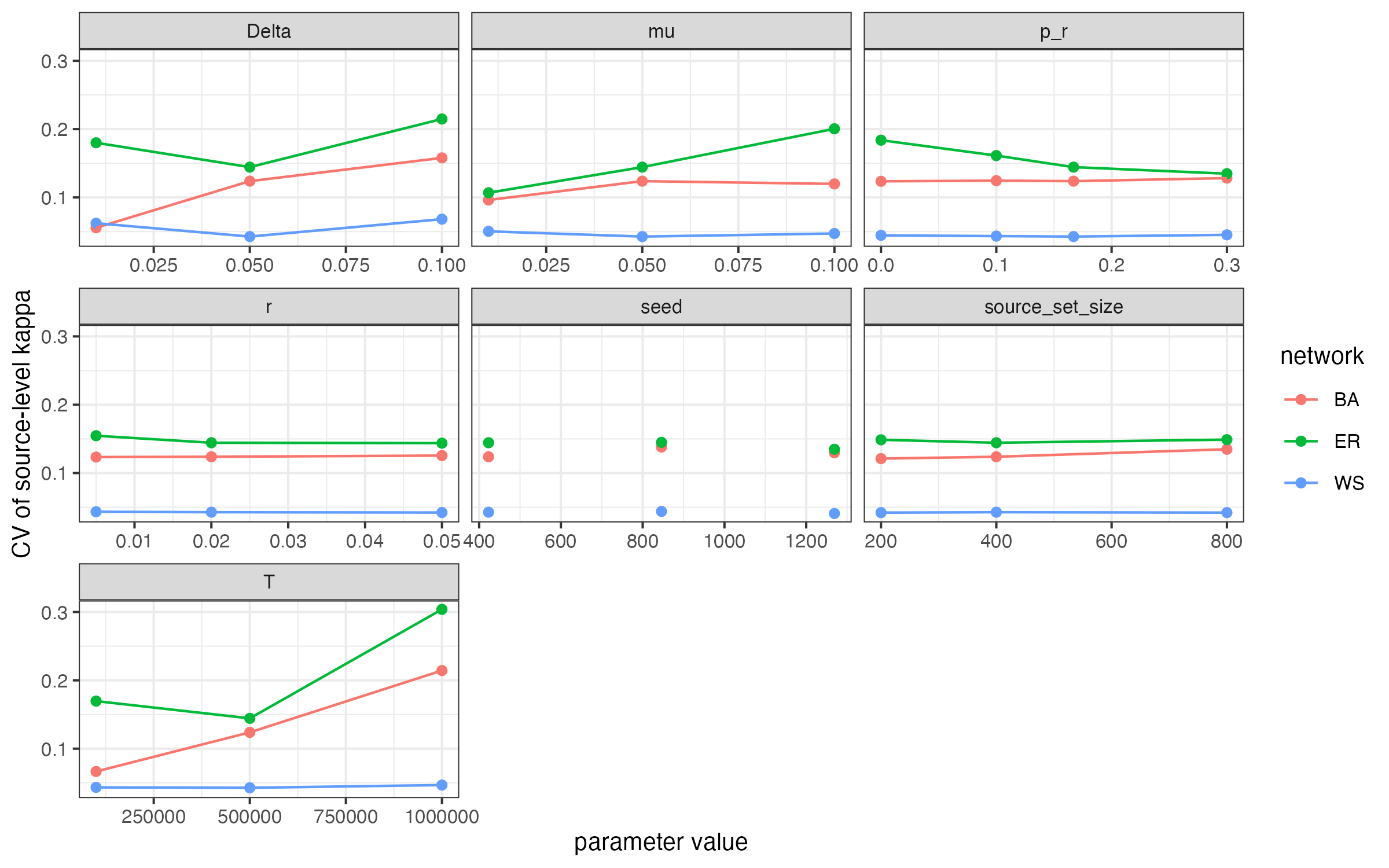}
\caption{Coefficient of variation of the source-specific decay rates, $\kappa_s$, for each parameter setting. The $T$ panel is ordered numerically. In the seed panel, points are shown for comparison only; no ordered trend is implied.}
\label{fig_sens_kappa_cv}
\end{figure}

The three aggregated CSV files used in this analysis are provided with this Supplementary Information.